\documentclass[aps,prd,superscriptaddress,nofootinbib,amsmath,amsfonts,preprintnumbers,groupedaddress,showpacs,10pt,english]{revtex4-1}
\usepackage{amsmath}
\usepackage{amssymb}
\usepackage{babel}
\usepackage{wrapfig}
\usepackage{cancel}
\usepackage{bigints}

\newcommand{\e}{\mathrm{e}}
\usepackage{relsize,exscale}
\makeatletter

\usepackage{array,multirow,graphicx}
\usepackage{dcolumn}
\usepackage{newlfont}
\usepackage{bm}
\usepackage[colorlinks,citecolor=blue,urlcolor=blue,linkcolor=blue]{hyperref}
\usepackage[figtopcap]{subfigure}
\usepackage{color}

\allowdisplaybreaks[4]

\begin{document}

\tolerance=5000

\title{%
Consistency between black hole and mimetic gravity \\
-- Case of $(2+1)$-dimensional gravity --}
\author{Shin'ichi~Nojiri}
\email{nojiri@gravity.phys.nagoya-u.ac.jp}
\affiliation{Department of Physics, Nagoya University, Nagoya 464-8602,
Japan \\
\& \\
Kobayashi-Maskawa Institute for the Origin of Particles and the Universe,
Nagoya University, Nagoya 464-8602, Japan }
\author{G.~G.~L.~Nashed}
\email{nashed@bue.edu.eg}
\affiliation {Centre for Theoretical Physics, The British University, P.O. Box
43, El Sherouk City, Cairo 11837, Egypt}


\date{}

\begin{abstract}

We show that the mimetic theory with the constraint $g^{\rho \sigma}\partial_\rho\phi \partial_\sigma\phi=1$ cannot
realize the black hole geometry with the horizon(s). To overcome such issue, we may change the mimetic constraint a little bit by
$\omega(\phi) g^{\rho \sigma}\partial_\rho\phi \partial_\sigma\phi=-1,$ where $\omega(\phi)$ is a function of the scalar field $\phi$.
As an example, we consider $(2+1)$-dimensional mimetic gravity with the mimetic potential and construct black hole (BH) solutions
by using this modified constraint.
We study three different classes: In the first class, we assume the Lagrange multiplier and mimetic potential are vanishing
and obtain a BH solution that fully matches the BH of GR despite the non-triviality of the mimetic field which ensures
the study presented in {\it  JCAP 01 (2019) 058}.
In the second class, we obtain a BH having constant mimetic potential and a non-trivial form of the Lagrange multiplier.
In the third class, we obtain a new BH solution with non-vanishing values of the mimetic field, the Lagrange multiplier, and the mimetic potential.
In any case, the solutions correspond to the space-time with only one horizon but we show that the formalism for the constraint works.

\end{abstract}

\pacs{04.50.Kd, 04.25.Nx, 04.40.Nr}
\keywords{Mimetic gravitational theory, analytic (2+1)-dimension BHs.}

\maketitle
\section{Introduction}\label{S1}

The study of a black hole (BH) is considered the most ravishing and astonishing astrophysical topic which appeared
as the solution of Einstein's general relativity (GR).
These topics yield a robust background for discovering different subjects of physics comprising thermodynamics,
quantum gravity, paramagnetism-ferromagnetism phase transition, superconducting phase transition, super-fluids,
spectroscopy, condensed matter physics, holographic hypothesis, information theory, and so on.
The discovery of the well-known confirmed correspondence between gravitational field in $N$-dimensional
anti-de Sitter (AdS) space-time and the conformal field theory (CFT), existing on the boundary of $(N-1)$-dimensional
space-time, which is known as $\mathrm{AdS}_N/\mathrm{CFT}_{N-1}$ correspondence or gauge/gravity duality, makes the study of BH solutions be a hot topic.
The understanding of the puzzle of information paradox that was studied in detail in \cite{Hawking:2016msc}
and the shadow of the super-massive BHs whose results given by $M87$ event horizon telescope \cite{EventHorizonTelescope:2019dse}
yield researchers to do more effort to understand the nature of BHs.
The existence of soft gravitons and/or soft photons, on the BH horizon and their quantum state information
which is stored on a holographic plate at the future boundary of the horizon, are discussed in \cite{Hawking:2016msc}.
The entropy and microscopic construction of BH near the horizon are discussed through the soft hairs
in \cite{Afshar:2016wfy, Haco:2018ske, Haco:2019ggi, Grumiller:2019fmp, Sheykhi:2020dkm}.

This paper is concerned with the study of the mimetic gravitational theory coupled with mimetic potential and the Lagrange multiplier.
The mimetic theory in $(3+1)$-dimensions is a new gravitational theory constructed recently as a new
prescription of the dark matter (DM) issue \cite{Chamseddine:2013kea,Chamseddine:2016uef}.
It is shown that the mimetic theory can have an additional longitudinal degree of
freedom to the gravitational field although the degree of freedom is not dynamical.
The additional longitudinal degree of freedom
is responsible for the mimetic DM in the case of $(3+1)$-dimensions.
Recently, it is discussed that an amendment of mimetic gravity in $(3+1)$-dimensions can solve the issue of the cosmological
singularities \cite{Chamseddine:2016uef} and the singularity at the center of a BH \cite{Chamseddine:2016ktu}.
Moreover, it is shown that the mimetic gravitational theory in $(3+1)$-dimensions forecasts
that gravitational wave (GW) can spread with a speed of light, confirming fully consistent with the results
of the event, GW170817, and its optical counterpart \cite{Casalino:2018tcd, Casalino:2018wnc}.
Moreover, it is shown that mimetic theory can investigate the flat rotation curves of spiral galaxies \cite{Vagnozzi:2017ilo, Sheykhi:2019gvk}.
Recently, a mimetic theory has gained a lot of attention in the domain of cosmology
\cite{Chamseddine:2014vna,Dutta:2017fjw,Abbassi:2018ywq,Zhong:2018tqn,Matsumoto:2016rsa,Nojiri:2014zqa,Odintsov:2015wwp,
Nojiri:2016vhu,Sadeghnezhad:2017hmr,Nashed:2021pkc,Gorji:2019ttx,Gorji:2018okn,Bouhmadi-Lopez:2017lbx,Gorji:2017cai,
Firouzjahi:2018xob,Sebastiani:2016ras,Chamseddine:2019bcn}
and in the domain of astrophysics \cite{Deruelle:2014zza,Myrzakulov:2015sea,Myrzakulov:2015kda,Astashenok:2015haa,Odintsov:2015cwa,
Nojiri:2017ygt,Odintsov:2018ggm,Oikonomou:2015lgy,Nashed:2014sea,Gorji:2020ten,Nashed:2018qag,Chen:2017ify,Nashed:2021ctg,Nashed2021,
Nashed:2021hgn,2018IJGMM..1550154N,Nashed:2018urj,BenAchour:2017ivq,Zheng:2017qfs,Elizalde:2020icc,Shen:2019nyp,Sheykhi:2020fqf}.

In the mimetic gravitational theory, the conformal degree of freedom of gravitational field can be isolated by introducing a relation
between the auxiliary metric $\bar{g}^{\alpha \beta}$, the physical metric $g_{\alpha \beta}$, and a mimetic field $\phi$
which is a scalar through the form,
\begin{align}
\label{trans1}
g_{\alpha\beta}=\mp \left(\bar{g}^{\mu \nu} \partial_\mu \phi\partial_\nu \phi \right) \bar{g}_{\alpha\beta}\,.
\end{align}
{ Eq.~(\ref{trans1}) has the scale invariance $\bar{g}_{\mu\nu}\to \e^\sigma \bar{g}_{\mu\nu}$ with a parameter $\sigma$.}
Equation~(\ref{trans1}) yields that the mimetic scalar field should give,
\begin{align}
\label{trans2}
g^{\alpha \beta}\partial_\alpha \phi \partial_\beta \phi= \mp 1\,.
\end{align}
{
In this paper, we choose the signature of the metric as $(+,-,-,-)$.
Therefore if we choose the negative sign in the r.h.s. of Eq.~(\ref{trans2}),
$\left( \partial_\alpha \phi \right)$ is a space-like vector and if we choose the positive sign,
$\left( \partial_\alpha \phi \right)$ is time-like.
When the cosmology is discussed, the positive sign has been often used and the mimetic matter can be interpreted as dust
but when we consider the static objects like black holes,
the negative sign has been used well.
}
Recently, many amendments of mimetic theory have been constructed like, mimetic gravity
with the Lagrange multipliers \cite{Cid:2012dh,Capozziello:2010uv} and $f(R)$ mimetic gravity.
Additionally, the existence of mimetic potential and the Lagrange multiplier assisted the
possibility for understanding different topics in the frame of cosmology, \cite{Odintsov:2015cwa}.
The mimetic theory of gravity is also extended to $f(R)$ mimetic theory of gravity \cite{Nojiri:2014zqa,Odintsov:2015wwp,
Oikonomou:2016pkp,Nashed:2010ocg,Oikonomou:2016fxb,Oikonomou:2015lgy,Nashed:2008ys,Myrzakulov:2016hrx,Odintsov:2015cwa,Odintsov:2016imq,
Odintsov:2016oyz,Nojiri:2017ygt,Odintsov:2018ggm,Chen:2020zzs} and the Gauss-Bonnet mimetic gravity \cite{Astashenok:2015haa,
Oikonomou:2015qha,Nashed:2007cu,Zhong:2016mfv,Zhong:2018tqn}.
More specific, a unified formulation of early inflation and late-time acceleration in the $f(R)$ mimetic gravity was formulated in \cite{Nojiri:2016vhu}
where confirmation of the inflationary era is established contrary to the $f(R)$ gravity.

In this paper, we show that the mimetic constraint (\ref{trans2}) is not consistent with the black hole geometry with the horizon(s).
In order to solve this problem, we modify the constraint (\ref{trans2}) by introducing a function $\omega(\phi)$ as follows,
\begin{align}
\label{trans3}
\omega(\phi) g^{\alpha \beta}\partial_\alpha \phi \partial_\beta \phi= \mp 1\,,
\end{align}
which is locally equivalent to  (\ref{trans2}) as we will see in the next section but $\omega(\phi)$ plays the important role when crossing the horizon.
{
We should also note that by introducing $\omega(\phi)$, we can treat the case that the time-like vector $\left( \partial_\alpha \phi \right)$ changes to
the space-like vector and vice versa in a solution even if we fix the negative or positive sign in the r.h.s. of Eq.~(\ref{trans3}).
}
We apply this formalism to $(2+1)$-dimensional mimetic gravity and show that the formalism works well.
We should also note that by the constraint (\ref{trans3}), Eq.~(\ref{trans1}) can be rewritten as
$g_{\alpha\beta}=\mp \omega(\phi) \left(\bar{g}^{\mu \nu} \partial_\mu \phi\partial_\nu \phi \right) \bar{g}_{\alpha\beta}$
and it is clear that there remains the scale invariance $\bar{g}_{\mu\nu}\to \e^\sigma \bar{g}_{\mu\nu}$.

The derivation of $(2+1)$-dimensional BH solutions of GR in AdS space-time, which is known
as the BTZ (Banados-Teitelboim-Zanelli) BHs \cite{Banados:1992wn}, is considered
as the most fascinating achievement in BH physics.
The metric of the BTZ black hole is given by
\begin{align}
\label{BTZ}
ds^2=-f(r) dt^2
+ \frac{1}{f(r)} dr^2 +r^2 \left(h(r)\,dt+d\theta\right)^2\,, \quad \textrm{where} \quad f(r)= \Lambda r^2 -M +\frac{J}{4r^2}\,, \textrm{and} \quad h(r)=-\frac{J}{2r^2}\,.
\end{align}
Here, $\Lambda$, $M$, and $J$ are the cosmological constant and the black hole's mass and angular momentum.
The $(2+1)$-dimensional BH solution of the Einstein GR gravitational theory yields an easy model to explain
and discuss some basic problems as quantum gravity, BH thermodynamics, holographic superconductors, string and
gauge/gravity duality, in the frame
of $\mathrm{AdS}_3 /\mathrm{CFT}_2$ \cite{Carlip:1995qv,Ashtekar:2002qc,Sarkar:2006tg,Witten:1998zw,Carlip:2005zn}.
The coincidence of the quasinormal modes in the $\mathrm{AdS}_3 /\mathrm{CFT}_2$
with the poles of the correlation function in the dual CFT gives quantitative proof for $\mathrm{AdS}_3 /\mathrm{CFT}_2$ \cite{Birmingham:2001pj}.
Moreover, the BTZ BHs are important tools to increase our knowledge of gravitational interaction in lower-dimensional
space-times \cite{Witten:2007kt}, especially, in the quantum gravity in $(2+1)$-dimension.
The geometries of the spinning $(2+1)$-dimensional BHs have been discussed in \cite{Banados:1992gq}.
The study of BTZ that has an electric charge, mass, and angular momentum has been analyzed in \cite{Martinez:1999qi, Fernando:1997rw}.
Furthermore, the $(2+1)$-dimensional BHs can supply a strong background to investigate one-dimensional
holographic superconductors \cite{Ren:2010ha,Liu:2011fy,KordZangeneh:2017zyy,Mohammadi:2018hxc,Ghotbabadi:2018ahu}.
There are many studies with different topics on $(2+1)$-dimensional BH solutions of gravitational field done in the literatures
(see e.g. \cite{Panotopoulos:2018pvu,Rincon:2018sgd,Clement:1995zt,Carlip:1995qv,Nojiri:1998ww,Emparan:1999fd,Cadoni:2008mw,
Parsons:2009si,Hendi:2012zz,Sheykhi:2014jia,Xu:2014xqa,Grumiller:2019tyl,Capozziello:2010uv,Sebastiani:2016ras}).

The arrangement of the present study is as follows:
In Section~\ref{S2}, we show that the mimetic constraint (\ref{trans2}) is inconsistent with the black hole geometry with the horizon(s)
and give a solution for this problem as given in (\ref{trans3}).
In Section~\ref{S3}, we apply the above formalism to the $(2+1)$-dimensional mimetic gravity coupled
with mimetic potential and the Lagrange multiplier.
In this section, we employ the field equations of mimetic theory to a $(2+1)$-dimensional space-time that has unequal metric potentials
and presents the non-linear differential equations which have five unknown functions, i.e., two of the metric potentials, mimetic scalar field,
mimetic potential field, and the Lagrange multiplier.
We solved this system of non-linear differential equations using three restrictions of the mimetic potential and the Lagrange multiplier
and obtain three different analytic BH solutions.
We summarize and discuss the results obtained in this study in the final Section~\ref{S66}.

\section{Inconsistency between mimetic gravity with black hole geometry}\label{S2}

The term ``mimetic DM" in $(3+1)$-dimensional space-time is delivered in the scientific society by Mukhanov and Chamseddine \cite{Chamseddine:2014vna}.

In this paper, we consider the black hole space-time but the mimetic constraint (\ref{trans2}) is not consistent with the black hole
space-time with horizon(s).
In (\ref{trans2}), we choose the minus sign as follows,
\begin{align}
\label{lambdavar0}
g^{\rho \sigma}\partial_\rho\phi \partial_\sigma\phi=-1\, .
\end{align}
Even if we choose the plus sign in (\ref{trans2}), the situation is not changed.
We consider the $D$ dimensional sphericaly symmetric and static space-time
with following line element,
\begin{align}
\label{metD}
ds^2=b(r)dt^2-\frac{dr^2}{b_1(r)}-r^2 d\Omega_{D-2}^2\, .
\end{align}
Here $d\Omega_{D-2}^2$ is the line element of $D-2$ dimensional unit sphere.
If we may also assume $\phi=\phi(r)$, the mimetic constraint (\ref{lambdavar0}) has the following form
\begin{align}
\label{cnstrnt1}
b_1 \left( \phi' \right)^2 = 1 \, .
\end{align}
The equation has no solution if $b_1$ is negative, $b_1<0$.
In the case of black hole geometry, $b$ vanishes and changes its signature at the horizon.
As we will see below, to avoid the curvature singularity, $b_1$ also must vanish at the horizon.
Therefore even if $b_1$ is positive, $b_1$ is negative in general, as in the Schwarzschild black hole or the BTZ black hole, which conflicts with Eq.~(\ref{cnstrnt1}).
This tells that the mimetic theory with the constraint (\ref{lambdavar0}) cannot realize the black hole geometry with the horizon(s).

In order to avoid the above problem, we may change the mimetic constraint in (\ref{lambdavar0}), a little bit as in (\ref{trans3}),
\begin{align}
\label{lambdavar1}
\omega(\phi) g^{\rho \sigma}\partial_\rho\phi \partial_\sigma\phi=-1\, .
\end{align}
Here $\omega(\phi)$ is a function of $\phi$.
If $\omega(\phi)$ is positive, we may define a new scalar field $\tilde\phi$ by $\tilde\phi = \int d\phi \sqrt{\omega(\phi)}$,
the constraint (\ref{lambdavar1}) is reduced to the form of (\ref{lambdavar0}),
\begin{align}
\label{lambdavar2}
g^{\rho \sigma}\partial_\rho \tilde\phi \partial_\sigma \tilde\phi=-1\, .
\end{align}
The signature of $\omega(\phi)$ can be, however, changed in general.
If we may also assume $\phi=\phi(r)$ and the space-time is given by (\ref{metD}),
instead of (\ref{cnstrnt1}), the constraint (\ref{lambdavar1}) has the following form,
\begin{align}
\label{cnstrnt2}
b_1 \omega(\phi) \left( \phi' \right)^2 = 1 \, .
\end{align}
Then for a solution of $\phi$ where $\omega(\phi)$ is positive when $b_1$ is positive and $\omega(\phi)$ is negative when $b_1$ is negative,
the constraint (\ref{cnstrnt2}) is consistent even inside the horizon.
As an example, we consider a simple case where
\begin{align}
\label{ex1}
\omega(\phi) = \frac{1}{\phi} \, .
\end{align}
Near the horizon, $b_1$ should behaves as
\begin{align}
\label{ex2}
b_1(r) \sim b_0 \left( r - r_\mathrm{h} \right) \, .
\end{align}
Here $r_\mathrm{h}$ is the radius of the horizon and $b_0$ is a positive constant.
Then a solution of (\ref{cnstrnt2}) with (\ref{ex1}) is given by
\begin{align}
\label{ex3}
\phi \sim \frac{r - r_\mathrm{h}}{b_0} \, .
\end{align}
Then the scalar field $\phi$ and therefore $\omega(\phi)$ change the sign at the horizon and
Eq.~ (\ref{cnstrnt2}) is consistent even inside the horizon.

In case there are several horizons, the problem might not be solved only by the choice in (\ref{ex1}).
A way to solve the problem in this case, we may choose
\begin{align}
\label{ex4}
\omega(\phi) = b_1 (\phi) \, .
\end{align}
In this case, the solution of (\ref{cnstrnt2}) is simply given by
\begin{align}
\label{ex5}
\phi = r \, .
\end{align}
Therefore it is clear that the problem is solved by the choice of (\ref{ex4}).
However, this choice may look rather artificial because it looks like we have assumed the solution from the beginning.
Anyway, the possibility of the choice (\ref{ex4}) tells that a model is realizing the solution of Eq.~(\ref{cnstrnt2}).

\section{$(2+1)$-dimensional BH solutions in mimetic gravity coupled with mimetic potential and Lagrange multiplier}\label{S3}

In this section, { we apply the formalism proposed in the last Section \ref{S2} to $(2+1)$-dimensional mimetic gravity and show that the formalism works well.}
The action of the mimetic gravitational theory that has the Lagrange multiplier $\lambda (\phi)$ and mimetic potential $V(\phi )$
has the following form \cite{Chamseddine:2014vna,Nojiri:2014zqa,Gorji:2020ten},
\begin{align}
\label{actionmimeticfraction}
S=\frac{1}{2\kappa^2}\int \mathrm{d}x^3\sqrt{-g}\left \{ R\left( g_{\mu \nu}\right)
-2\Lambda-V(\phi)+\lambda \left(\omega(\phi) g^{\mu \nu}\partial_\mu\phi\partial_\nu\phi
+1\right)\right \}+S_\mathrm{matt}\, ,
\end{align}
where $\kappa^2$ is the Einstein gravitational constant which in the relativistic units equals $\kappa^2=8\pi$ and $S_\mathrm{matt}$
is the action of the matter fields.
Furthermore, we assume the auxiliary scalar mimetic field, $\phi$, relies on the radial coordinate $r$.
The variation of the action (\ref{actionmimeticfraction}) w.r.t. the metric tensor $g_{\mu \nu}$, yield the field equations,
\begin{align}
\label{aeden}
0=R_{\mu \nu} -\frac{1}{2} g_{\mu \nu}\left(R-2\Lambda\right)
+\frac{1}{2}g_{\mu \nu}\left\{\lambda \left( \omega(\phi) g^{\rho\sigma}\partial_\rho\phi\partial_\sigma\phi+1\right) -V(\phi)\right\}
 -\lambda \omega(\phi)\partial_\mu\phi \partial_\nu\phi +8\pi T_{\mu \nu} \, ,
\end{align}
with $T_{\mu \nu}$ being the energy-momentum tensor of the matter fluids.
In the present study, we only consider the case $T_{\mu \nu}=0$.
Moreover, the variation of the action w.r.t. the auxiliary mimetic scalar field $\phi$, gives
\begin{align}
\label{scalvar}
 2\nabla^\mu (\lambda\omega(\phi) \partial_\mu\phi) -\lambda \omega'(\phi) g^{\rho\sigma}\partial_\rho\phi\partial_\sigma\phi+V'(\phi)=0\, ,
\end{align}
where the ``prime'' means the differentiation w.r.t. the auxiliary scalar $\phi$.
Finally, the variation of the action (\ref{actionmimeticfraction}) w.r.t. the Lagrange multiplier $\lambda$, gives
\begin{align}
\label{lambdavar}
\omega(\phi) g^{\rho \sigma}\partial_\rho\phi \partial_\sigma\phi=-1\, ,
\end{align}
which coincides with Eq.~(\ref{lambdavar1}).
We should note, the scalar field equation (\ref{scalvar}) can be obtained from (\ref{aeden}) and (\ref{lambdavar}) when $T_{\mu\nu}=0$ or
by using the conservation law.
Therefore we do not use (\ref{scalvar}) hereafter.

Now, we are ready to apply the field equations of the mimetic theory, (\ref{aeden}), and (\ref{scalvar}) to a $(2+1)$-dimensional space-time
that has two unknown functions and takes the form in the coordinates $(t,r,\theta)$ as,
\begin{align}
\label{met}
ds^2=b(r)dt^2-\frac{dr^2}{b_1(r)}-r^2\,d\theta^2\,.
\end{align}
Here $b(r)$ and $b_1(r)$ are unknown functions of radial coordinate $r$, that will fixed through the use of the field equations (\ref{aeden}), and (\ref{scalvar}).
Using Eq.~(\ref{met}), we get the Ricci scalar as,
\begin{align}
\label{Ricci}
R={\frac{b' b'_1 b r+2b_1b'' b r-b'^2b_1 r+2b_1b'b +2 b'_1b^2}{2b^2r}}\,.
\end{align}
The expression (\ref{Ricci}) tells that the scalar curvature $R$ might diverge when $b=0$ because there is $b^2$ in the denominator.
If $R$ diverges, the radius where $b$ vanishes does not correspond to the horizon but the singularity.
To check if the radius with $b=0$ corresponds to the horizon or the singularity, we assume, as in the standard Schwarzschild space-time
$b_1$ is written in the form $b_1 = b_2 b$, where $b_2$ is finite when $b=0$.
Then the scalar curvature $R$ in (\ref{Ricci}) is rewritten as
\begin{align}
\label{Ricci2}
R= \frac{b' b'_2 r+2b_2b''r + 2 b'_2 b + 4b_2 b' }{2 r}\,.
\end{align}
Therefore if $b_2$ vanishes when $b$ vanishes, the scalar curvature $R$ is finite and the radius where $b$ vanishes corresponds to the
horizon.

Applying the field equations (\ref{aeden}) and (\ref{scalvar}) to space-time (\ref{met}), after using Eq.~(\ref{Ricci}),
we obtain the following non-linear differential equations:

\

\noindent
The $(t,t)$-component of the field equation (\ref{aeden}) is
\begin{align}
\label{tt}
0=\frac{b'_1+Vr-2r\Lambda}{2 r}\,,
\end{align}
the $(r,r)$-component is
\begin{align}
\label{rr}
0=\frac{b_1b' -2\lambda bb_1r\omega\phi'^2 - 2br\Lambda+V\,b r}{2 rb}\,,
\end{align}
the $(\theta,\theta)$-component is
\begin{align}
\label{thth}
0=\frac{bb'_1b'+2b''bb_1-b'^2b_1 - 4b^2\Lambda+2V\,b^2}{4b^2}\, ,
\end{align}
where $b\equiv b(r)$, $b_1\equiv b_1(r)$, $V\equiv V(r)$ \footnote{Due to the nature of our study in this research, we assume the mimetic potential and
the Lagrange multiplier to depend on the radial coordinate $r$.}, $\lambda \equiv \lambda(r)$, $b'=\frac{db}{dr}$, $b'_1=\frac{db_1}{dr}$, $V'=\frac{dV}{dr}$,
$\phi'=\frac{d\phi}{dr}$, and $\lambda'=\frac{d\lambda}{dr}$.
The above non-linear differential equations, i.e., (\ref{tt}), (\ref{rr}), and (\ref{thth})
are going to be solved for the following three different classes:

\

\noindent
\underline{Case I: $V(r)=\lambda(r)=0$.}\vspace{0.2cm}\\
When $V(r)=\lambda(r)=0$, the analytic solution of the non-linear differential equations (\ref{tt}), (\ref{rr}), and (\ref{thth}) 
after resealing the constants of integration, takes the following form,
\begin{align}
\label{sol1}
b(r)= b_1(r)=\Lambda r^2+C=\Lambda r^2-M\,, \quad C=-M\, .
\end{align}
If we choose $\omega(\phi)$ as in (\ref{ex1}), we obtain
\begin{align}
\label{sol1phi}
\phi(r)= \frac{1}{2}\left\{ \frac{1}{\sqrt{\Lambda}}\ln\left(\sqrt{\frac{\Lambda r^2}{M}}+\sqrt{\frac{\Lambda r^2}{M} - 1}\right) \right\}^2 \,.
\end{align}
Because $b=b_1$, the arguments around Eq.~(\ref{Ricci2}) tell that the radius where $b(r)=b_1(r)=0$, that is, $r=\sqrt{\frac{M}{\Lambda}}$
when $\frac{M}{\Lambda}>0$ corresponds to the horizon radius.
Inside the horizon, we find $\sqrt{\frac{\Lambda r^2}{M} - 1}=\pm i \sqrt{1-\frac{\Lambda r^2}{M}}$.
Because $\frac{\Lambda r^2}{M}<1$ inside the horizon and $\left( \sqrt{\frac{\Lambda r^2}{M}} \right)^2 + \left( \sqrt{1-\frac{\Lambda r^2}{M}} \right)^2=1$,
we may write $\sqrt{\frac{\Lambda r^2}{M}} = \cos\theta$ and we find
\begin{align}
\label{sol1phi2},
\phi(r)= - \frac{\theta^2}{2\Lambda} \,,
\end{align}
which is surely real and negative.
The expression in (\ref{sol1phi}) tells $\phi$ is positive outside the horizon.
Therefore due to the factor $\omega(\phi)$, we can connect the region inside the horizon with the region outside the horizon smoothly and
$\phi$ vanishes on the horizon.
Equation~(\ref{sol1}) shows that when the mimetic potential and the Lagrange multiplier are vanishing, we return to the well-known BTZ BH
of GR solution despite the non-vanishing of the mimetic field.
This result is consistent with the previous studies \cite{Nashed:2018qag}.
It is of interest to note that the mimetic field of Eq.~(\ref{sol1phi}) satisfies the constraints (\ref{cnstrnt2}).
The relevant physics of the BTZ space-time is well studied in the literature.
From now on, we will put the cosmological constant equal
to zero, i.e., $\Lambda=0$ to show the effect of the Lagrange multiplier and the mimetic potential on the geometry under consideration.

\

\noindent
\underline{Case II: $V(r)=\mathrm{const}=c_1$.}\\
By choosing $V(r)=\mathrm{const}=c_1$ and using $\omega(\phi)$ in (\ref{ex1}), we find
\begin{align}
\label{sol2}
b_1=&\, c_2-\frac{c_3r^2}2\,, \quad \lambda(r)=\frac{c_2}{r \left( c_2c_3\sqrt{c_1r^2-2c_2}-r \right)}
\,,\nonumber \\
b(r)=& c_4\e^{
\int \frac{2 \left(2 c_2-c_1r^2+c_1c_2c_3r
\sqrt{c_1r^2-2 c_2} \right)}{ \left(c_2c_3\sqrt{c_1r^2-2c_2}-r\right) \left( c_1r^2-2c_2 \right) }dr} \nonumber\\
=& -\left\{c_4 \left(r{c_3}^2{c_2}^2c_1 -r -\sqrt2c_3 {c_2}^\frac{3}{2} C_0\right) \left( 2 {c_3}^2{c_2}^3c_1+c_1r\sqrt2c_3 {c_2}^\frac{3}{2}C_0
 -2 c_2-\sqrt2C_0\sqrt{c_2}\sqrt{c_1 r^2-2 c_2} \right) \right. \nonumber\\
& \left. \times \left({c_3}^2{c_2}^2c_1 r^2 - r^2-2 {c_3}^2{c_2}^3 \right) \right\} \left[\left(c_1r\sqrt2c_3 {c_2}^\frac{3}{2}C_0
 -2 {c_3}^2{c_2}^3c_1 +2 c_2 +\sqrt2C_0\sqrt{c_2}\sqrt{c_1 r^2-2 c_2} \right) \right. \nonumber\\
& \left. \times \left( r{c_3}^2{c_2}^2c_1 -r +\sqrt2c_3 {c_2}^\frac{3}{2}C_0 \right)\right]^{-1} \, , \nonumber \\
\phi(r)=& \frac{1}{\omega(r)} = \frac{1}{c_1} \left\{ \tan^{-1}\left(\frac{\sqrt{c_1}r}{\sqrt{2c_2-c_1r^2}}\right) \right\}^2
\,,
\end{align}
where we have put $C_0=\sqrt{{c_3}^2{c_2}^2c_1-1}$.
As Eq.~(\ref{sol2}) shows that the constant $c_1$ cannot have a zero value which means that it cannot reduce to the first case.
In order that the radius where $b_1=0$, that is, $r=\sqrt{\frac{2 c_2}{c_3}}$ when $\frac{2 c_2}{c_3}>0$, is the horizon radius, by the arguments around Eq.~(\ref{Ricci2}), $b$
should also vanish when $b$ vanishes, which gives a constraint on the parameters, as follows,
\begin{align}
\label{vanishngb}
0= \left(c_3{c_2}^2 c_1 - 1 -{c_3}^\frac{1}{2} c_2 C_0\right)
\left( {c_3}^2{c_2}^2c_1 +c_1 {c_3}^\frac{1}{2} c_2 C_0 - 1 - C_0 \sqrt{\frac{c_1}{c_3} - 1 } \right) \left({c_3}^2 {c_2}^2c_1 - 1 - {c_3}^3{c_2}^2 \right) \, .
\end{align}
Here we have assumed $c_2$ and $c_4$ do not vanish.
Other radii where $b$ vanishes corresponding to the curvature singularity and they are not the horizon radius.
We may choose the parameters so that the singular surfaces do not appear or the singular surfaces are inside of the black hole horizon.
In order that the black hole exists, the radius $r=\sqrt{\frac{2 c_2}{c_3}}$ to be the black hole horizon, which require $c_2$ and $c_3$ are negative.
Anyway, this solution can have only one horizon as in the BTZ black hole with the angular momentum $J=0$.

The expression of $\phi(r)$ is valid when $r<\sqrt{\frac{2 c_2}{c_3}}$.
When $r>\sqrt{\frac{2 c_2}{c_3}}$, because
$\tan^{-1}\left(\frac{\sqrt{c_1}r}{\sqrt{2c_2-c_1r^2}}\right) = \pm i \tanh^{-1}\left(\frac{\sqrt{c_1}r}{\sqrt{c_1r^2- 2c_2}}\right)$,
we find $\phi(r)$ becomes negative and real
\begin{align}
\label{phil}
\phi(r)= \frac{1}{\omega(r)} = - \frac{1}{c_1} \left\{ \tanh^{-1}\left(\frac{\sqrt{c_1}r}{\sqrt{c_1r^2- 2c_2}}\right) \right\}^2 \,.
\end{align}
Therefore the formalism using $\omega(\phi)$ in (\ref{lambdavar1}) is valid even in Case II.

\

\noindent
\underline{Case III: $\lambda(r)\neq0$ and $V(r)\neq 0$.}\\

The differential equations~(\ref{tt}), (\ref{rr}), and (\ref{thth}) possess many analytic solutions, however, in this study, we are going to consider
only the physical one which yields metric potential $b$ asymptotes as AdS/dS and also the mimetic potential behave asymptotically
for $r\to \infty$ as constant as it should for any physical model.
Now we are going to use the constraint (\ref{cnstrnt2}) then we have three differential equations, ~(\ref{tt}), (\ref{rr}), and (\ref{thth}),
in four unknowns $b(r)$, $b_1(r)$, $\lambda(r)$ and $V(r)$.
Therefore, we have to fix anyone of the previous unknowns by hand. Here we will use the form of $V(r)$ as,
\begin{align}
\label{sol333}
V(r) =&\, c_5+\frac{c_6} r +\frac{c_7}{r^3}\,,
\end{align}
where $c_5, c_6$, and $c_7$ are constants. The merit of the form of Eq. (\ref{sol333}) is that when $r\to \infty$ gives
constant potential as it should be from the view point of physics. Using Eq. (\ref{sol333}) in Eqs.~(\ref{tt}), (\ref{rr}),
and (\ref{thth}) we get the solution of the remanning unknowns in the form,
\begin{align}
\label{sol3}
b(r) =&\, \frac{r^2}{2} \left\{ \int \frac{c_9dr}{\sqrt{r^3 \left( 2c_7 c_8 - r - 2c_6 c_8 r^2 - c_5 c_8 r^3\right) }} +c_{10} \right\}^2
\,,\nonumber\\
b_1(r) =& \frac{2c_7 c_8 - r - 2c_6 c_8 r^2 - c_5 c_8 r^3}{2rc_8}\,,
 \nonumber\\
\lambda(r)=&\, \frac{1}{2} \left\{ -c_9 \e^{2 \int \frac{\left( 2 + 5 c_6 c_8 r + 3 c_5 c_8 r^2\right) dr}{2c_7 c_8 - r - 2c_6 c_8 r^2 - c_5 c_8 r^3}}
\frac{\left( c_6 c_8 r^2 -3 c_7 c_8 +r \right)}{\sqrt{r^3 \left( 2c_7 c_8 - r - 2c_6 c_8 r^2 - c_5 c_8 r^3\right) }}
 -c_{10} \left( -3 c_7 c_8 + r + c_6 c_8 r^2 \right) \right. \nonumber\\
& \left. - \frac{rc_9\left( -2 c_7 c_8 + r +2 c_6 c_8 r^2 + c_5 c_8 r^3 \right)}{\sqrt{r^3 \left( 2c_7 c_8 - r - 2c_6 c_8 r^2 - c_5 c_8 r^3\right) }} \right\}
\left( \frac{1}{\sqrt{r^3 \left( 2c_7 c_8 - r - 2c_6 c_8 r^2 - c_5 c_8 r^3\right) }} + c_{10} \right)
r^{-3} {c_8}^{-1} \, .
\end{align}
As discussed in the paragraph including Equations~(\ref{Ricci}) and (\ref{Ricci2}), $b_1(r)$ must vanish when $b(r)$ vanishes.
First, let us consider the case that $b_1(r)$ in (\ref{sol3}) has only one zero $b_1(r)=0$ at $r=r_\mathrm{h}$.
We can always adjust the constant of the integration $c_{10}$ so that $b(r)$ also vanishes at $r=r_\mathrm{h}$, $b\left(r_\mathrm{h}\right)=0$.
Then, in this case, the radius $r_\mathrm{h}$ can be regarded as the horizon radius.
We may also consider the case that $b_1(r)$ has two zeros at $r=r_\mathrm{h}$ and $r=r_{\mathrm{h}1}$ and we assume $r_\mathrm{h}>r_{\mathrm{h}1}$.
In this case, $b_1(r)$ can be rewritten as follows,
\begin{align}
\label{3b1}
b_1(r) =& \frac{- c_5 \left( r - r_\mathrm{h} \right)\left( r - r_{\mathrm{h}1}\right) \left( r + r_0 \right)}{2r}\, .
\end{align}
Here $r_0$ is a positive constant.
We should note the constant $c_5$ must be negative.
As in the above case that $b_1(r)$ has only one zero, by choosing $c_{10}$ so that $b(r)$ also vanishes at $r=r_\mathrm{h}$, we find
\begin{align}
\label{3b}
b(r) = \frac{r^2}{2} \left\{ \int_{r_\mathrm{h}}^r \frac{c_9dr}
{\sqrt{ - c_5r^3\left( r - r_\mathrm{h} \right)\left( r - r_{\mathrm{h}1}\right) \left( r + r_0 \right)}} \right\}^2
= \frac{r^2}{4} \left\{ \int_{r_\mathrm{h}}^r \frac{c_9dr}{\sqrt{ r b_1(r) }} \right\}^2 \,.
\end{align}
When $r_{\mathrm{h}1}<r<r_\mathrm{h}$, $b_1(r)$ is negative and therefore $b(r)$ becomes negative,
\begin{align}
\label{3b2}
b(r) = - \frac{r^2}{4} \left\{ \int_{r_\mathrm{h}}^r \frac{c_9dr}{\sqrt{- r b_1(r) }} \right\}^2 \,.
\end{align}
In the region $r<r_{\mathrm{h}1}$, however, $b(r)$ becomes a complex number,
\begin{align}
\label{3b3}
b(r) = \frac{r^2}{4} \left\{ \pm i \int_{r_\mathrm{h}}^{r_{\mathrm{h}1}} \frac{c_9dr}{\sqrt{- r b_1(r) }}
+ \int_{r_{\mathrm{h}1}}^r \frac{c_9dr}{\sqrt{ r b_1(r) }} \right\}^2 \,.
\end{align}
Therefore the model becomes physically inconsistent and irrelevant or
there is the end of space at $r=r_{\mathrm{h}1}$.
The above arguments show that the model can have only one horizon, even in Case III as in Cases I and II.

Using Eq.~(\ref{sol3}), we get the mimetic field $\phi$ and the function $\omega$ in the form,
\begin{align}
\label{scal3}
\phi(r)=\frac{1}{\omega(r)}=\left( \int \!{\frac { dr }{\sqrt {4\,{\frac {c_7} r}-4\,rc_6
-2\, r^2 c_5-2\,{c_8}^{-1}}}} \right) ^2 +2\,\int
\!{\frac { dr }{\sqrt {4\,{\frac {c_7} r}-4\,rc_6-2\, r^2 c_5-2\,{c_8}^{-1}}}}c_9+{c_9}^2 \,.
\end{align}
We choose the constant of the integration $c_9$ so that $\phi$ vanishes when $b(r)$ and $b_1(r)$ vanishes, that is
\begin{align}
\label{scal3b}
\phi(r)=\left( \int_{r_\mathrm{h}}^r \frac{dr}{\sqrt {\frac {4c_7}{r} -4 rc_6 -2 r^2 c_5 -2{c_8}^{-1}}} \right)^2
=\frac{1}{2}\left( \int_{r_\mathrm{h}}^r \frac{dr}{\sqrt {b_1(r)}} \right)^2 \, .
\end{align}
The expression (\ref{scal3b}) is valid for $r$ where $b_1(r)>0$.
For $r$ where $b_1(r)<0$, $\phi$ becomes negative and real again,
\begin{align}
\label{scal3c}
\phi(r)
= - \frac{1}{2}\left( \int_{r_\mathrm{h}}^r \frac{dr}{\sqrt {- b_1(r)}} \right)^2 \, .
\end{align}
Therefore the formalism using $\omega(\phi)$ in  (\ref{lambdavar1}) is valid, again.

\section{Summary and Discussion}\label{S66}

The topic of DM became a hot topic based on the data coming from observations of astrophysics as well as cosmology.
To explain the issue of dark matter, scientists supposed two directions:
The first one is to amend the field equation of GR and the second direction is to amend the standard model by building a new particle kind.
Many studies have explained that the two directions are equal \cite{Sebastiani:2016ras}.
Additionally, it is familiar in the frame of amended theories of gravity that they have new degrees of freedom in addition to the usual massless graviton of GR.
As we discussed in Section~\ref{S1} that the target of the mimetic theory of gravity was mimetic DM, and was a good theory to explain the presence of cold DM.
We have shown, however, that the mimetic theory with the constraint $g^{\rho \sigma}\partial_\rho\phi \partial_\sigma\phi=-1$
cannot realize the black hole geometry with the horizon(s).
To bypass such problem, we changed the mimetic constraint as $\omega(\phi) g^{\rho \sigma}\partial_\rho\phi \partial_\sigma\phi=-1,$
where $\omega(\phi)$ is a function of the scalar field $\phi$.
Because we can choose $\omega(\phi)$ changes its signature at the horizon, the constraint becomes consistent with the existence of the horizon(s).
As applications of the formalism of the constraint, we considered the mimetic gravity in $(2+1)$-dimensional space-time and found that the
formalism works well.

{
In future work, we may consider the time-dependent black hole solution or the black hole solution in the expanding universe, which is difficult to be found
even in other gravity theories.
In the case of the mimetic theory, the vector $(\partial_\alpha \phi)$ is time-like when we consider the cosmology and
the vector is space-like when we consider the static object like a black hole.
If we use the original mimetic constraint given by Eq.  (\ref{trans2}), it is difficult to include both the time-like vector and the space-like vector for $(\partial_\alpha \phi)$
but by introducing $\omega(\phi)$ as in Eq. (\ref{trans3}), it becomes possible to include both the time-like and the space-like vectors in a solution,
which may help to construct the time-dependent black hole solution or the black hole solution in the expanding universe.
}

\begin{acknowledgments}
This work is supported by the JSPS Grant-in-Aid for Scientific Research (C)
No. 18K03615 (S.N.).
\end{acknowledgments}

%

\end{document}